# High thermoelectric performance of graphite nanofibers

**Van-Truong Tran[1][*], Jérôme Saint-Martin[2], Philippe Dollfus[2] and Sebastian Volz[1, 3]**

[1]EM2C, CentraleSupélec, Université Paris Saclay, CNRS, 92295 Châtenay Malabry, France
[2]C2N, CNRS, Université Paris-sud, Université Paris Saclay, 91405 Orsay, France
[3]LIMMS, Institute of Industrial Science, University of Tokyo, CNRS-IIS UMI2820, 4-6-1 Komaba Meguro-Ku, Tokyo 153-8505, Japan
*vantruongtran.nanophys@gmail.com

**Abstract**

Graphite nanofibers (GNFs) have been demonstrated to be a promising material for hydrogen storage and heat management in electronic devices. Here, by means of first-principles and transport simulations, we show that GNFs can also be an excellent material for thermoelectric applications thanks to the interlayer weak van der Waals interaction that induces low thermal conductance and a step-like shape in the electronic transmission with mini-gaps, which are necessary ingredients to achieve high thermoelectric performance. This study unveils that the platelet form of GNFs in which graphite layers are perpendicular to the fiber axis can exhibit outstanding thermoelectric properties with a figure of merit *ZT* reaching 3.55 in a 0.5 nm diameter fiber and 1.1 for a 1.1 nm diameter one. Interestingly, by introducing [14]C isotope doping, *ZT* can even be enhanced up to more than 5, and more than 8 if we include the effect of finite phonon mean-free path, which demonstrates the amazing thermoelectric potential of GNFs.

**Introduction**

Carbon fibers have been recognized since the 1950s as a tremendously important material in aerospace, construction, and automobile industries as it can be used to reinforce the mechanical properties of plastics, metals and ceramics, and to synthesize strong composite materials.[1] For these kinds of application, carbon fibers are commonly produced in the fiber form of few micro meters in diameter and length. [1]

Since the 1980s, carbon fibers have been fabricated with diameters less than 100 nm and took the name of carbon nanofibers.[2] Among different conformations of carbon nanofibers such as carbon nanotubes or "cup-stacked" Carbon nanofibers, graphite nanofibers (GNFs) are the most natural ones and the cheapest to produce. [2–4]

Basically, graphite and other carbon-based nanofibers are commonly synthesized using chemical vapor deposition (CVD) techniques. Fibers having a diameter of 2 nm to 100 nm and a length of a few nm to 100 µm can be achieved. [2,3,5,6] Depending on the arrangement of graphite layers with respect to the fiber axis, GNFs can be in the form of either ribbons, platelet or herringbone fibers. [2,7]

By intercalating hydrogen atoms in graphite layers, it has been demonstrated that GNFs can be efficiently used for hydrogen storage with a view to developing fuel cell applications.[2,7] Other works have shown that the thermal performance of phase change materials can be improved by introducing GNFs that allows reducing the maximum system temperature, which opens a route for the heat management of electronic devices. [4]

The latter effect of GNFs, in fact, comes from the anisotopic thermal conductivity of graphite. [8] It is experimentally demonstrated that thermal transport along the c-axis (perpendicular to graphite layers) is significantly weaker than along the parallel directions. [8] From the viewpoint of thermoelectric applications, this property is particularly interesting as it suggests that low thermal conductance along the c-axis might be exploited to achieve high thermoelectric figure of merit $ZT$. Actually, some previous experimental [9] and theoretical works [10,11] have shown that thermal transport across few layers of graphene or other 2D materials is indeed remarkably smaller than that in the in-plane directions of graphene sheets, leading to enhanced thermoelectric properties. However, sufficient attention has not been paid to the investigation of thermoelectric transport along only the c-axis (i.e. along the direction perpendicular to graphene planes) in graphite and graphite fibers.

Recently, Ma et al.[12] have experimentally examined the thermoelectric properties of graphite fibers oriented along the c-axis. Although this work has pointed out that the figure of merit $ZT$ significantly increases as the temperature increases, the obtained values of $ZT$ were still extremely small ($ZT$~$10^{-6}$) and thus not relevant for thermoelectric generation. The low thermoelectric performance of this graphite fiber is due to the large cross-section (macroscopic scale) which results in both a low Seebeck coefficient and a high thermal conductance.

As pointed out by Hicks and Dresselhaus in 1993 [13,14], nanoscale low-dimensional materials should offer a better thermoelectric performance than their bulk counterparts thanks to quantization and interface effects. Indeed, these two combined effects should induce an enhanced Seebeck effect and a reduced thermal transfer.

In this work, we theoretically explore for the first time the thermoelectric potentials of GNFs. By using ab initio calculations for electrons and a semi-empirical Force Constant (FC) model for phonons, in combination with Green's function formalism of transport, we demonstrate that very high thermoelectric performance can be achieved with *ZT* reaching 3.55 in perfect GNFs 0.5 nm in diameter and even up to 5.07 when introducing $^{14}$C isotope doping. The impact of the inelastic phonon scattering is also discussed and shown to further enhance the thermoelectric performance. Those outcomes not only open a path for new applications of GNFs but also put GNFs in the list of the most promising candidates for nanoscale thermoelectric applications.

**Results and discussions**

In figure 1, the sketch of the studied structure is shown with the indication of the transport direction along the c-axis, perpendicular to the graphite layers. This type of graphite nanofiber is commonly referred to as "platelet" GNFs.[2,4] Each unit cell contains two sub-layers and the diameter of the fiber is characterized by the number of slices in each sub-layer along the armchair and zigzag edges, $M_A$ and $M_Z$, respectively, or by the corresponding length $d_A$ and $d_Z$ with $d_A = (3M_A - 2) \times a_0 / 2$ and $d_Z = (M_Z - 1) \times \sqrt{3}/2 \times a_0$, where $a_0 \approx 0.142$ nm is the distance between two nearest in-plane carbon atoms. For example, in figure 1(c) the structure has a cross-section characterized by $M_A = 4$ ($d_A \approx 0.71$ nm) and $M_Z = 5$ ($d_Z \approx 0.492$ nm).

It is worth to note that since the 1980s, the CVD method of GNF growth has been improved, making it possible to synthesize smaller and higher quality structures. As recently reported, the smallest graphene ribbon of width 5 dimer lines (~0.5 nm) has been grown successfully in ultra-high vacuum. [15] It is thus expected that growing GNFs of diameter less than 2 nm, possibly around 1 nm, is accessible using current techniques. In this article, keeping in mind the idea to focus on structures that can both be potentially fabricated in the short term and studied theoretically by means of first-principle methods, we considered GNFs of diameter ranging between 0.5 nm and 1.1 nm.

## The electronic and phononic properties of GNFs

In figure 2, we show the electronic (upper row) and the phononic (lower row) properties of a GNFs having a cross-section size [ $M_A = 4, M_Z = 5$ ].

The electronic quantities were obtained from ab initio calculations with the SIESTA code.[16] The band structure and the transmission were calculated from the relaxed structure and the transmission was used to compute other transport properties such as the electrical conductance $G_e$, the Seebeck coefficient $S$ and the electron thermal conductance $K_e$, as detailed in the Supporting Information. As it can be seen from figure 2(a), the band structure contains several mini-gaps within the conduction and valence bands, which result in the step-like shape of the transmission profile as depicted in figure 2(b). As demonstrated in ref [17], the step-like form of the transmission is an ideal configuration to achieve high power factor and thermoelectric performance. The nearly flat band at zero energy in figure 2(a) is the origin of the sharp peak at zero energy in the electron transmission $T_e$ in figure 2(b). The band that crosses this nearly flat band with high group velocity can induce higher electron transport around the zero energy level. It should be noted that this band actually originates from the long range of van der Waals (vdW) interactions between atoms of different graphite layers. As such, it is not found from common tight-binding calculations limited to first-nearest layer interactions. As a result, this band induces high electrical conductance $G_e$ around the zero energy level as displayed in figure 2(c). The highest peak of the Seebeck coefficient occurs far away from the zero energy point, close to the regions of large mini-gaps (dashed-blue line in figure 2(c)).

In figure 2(d) we show the phonon dispersion of the GNF obtained from the FC model. The details of this calculation are given in the Supporting Information. It can be observed in the phonon dispersion that all dispersive (high-velocity) bands are in the frequency-range below 200 cm$^{-1}$, while the upper phonon bands are almost flat and thus weakly conductive. This result is similar to that obtained in bulk graphite along the c-axis as demonstrated in refs. [18,19] and also in the Supporting Information. As a consequence, the phonon transmission $T_p$ is significant only in this low-frequency range as can be observed from the solid-black line in figure 2(e). By comparing the dispersion of graphite along the c-axis with that along other axes,[18] we can understand that this behaviour is due to the weakness of the vdW interactions between atoms of distinct graphite layers compared to the strength of the in-plane interactions.

To verify the weakness of the vdW interactions in our GNF structures, we also computed the phonon transmission for in-plane armchair and zigzag graphene nanoribbons (AGNRs and ZGNRs) of same width as the GNF, i.e. 5 dimer lines for the AGNR and 4 chain lines for the ZGNR. The results are given in the inset of figure 2(e) clearly showing that the phonon transmission of GNRs spreads over the full-frequency range from 0 up to about 1600 cm$^{-1}$, i.e. in a much larger range compared to the one of the transmission along the c-axis of the GNF. It is thus not surprising that the phonon conductance of these two GNRs is much higher than that of the GNF, as shown in figure 2(f). At room temperature, the phonon conductance obtained in the GNF is about 0.098 nW/K, which is 8.4 times smaller than that in the AGNR and 15 times smaller than that in the ZGNR, i.e 0.82 nW/K and 1.47 nW/K, respectively.

The low phonon conductance is often the most important factor to achieve high thermoelectric performance. Thus, the results above suggest that the thermoelectric capacity of GNFs should be much better than that of graphene nano-ribbons.

*Thermoelectric capacity of GNFs*

In figure 3(a), the figure of merit *ZT* is plotted at two different temperatures for the GNF of cross-section size [$M_A = 4, M_Z = 5$]. At 300 K the maximum value $ZT_{max} = 2.85$ is obtained at the chemical energy $\mu = -1.18$ eV. This value is actually much higher than the one of about 0.35 that has been found in graphene ribbon structures of similar size at the same temperature. [20] The red line shows that *ZT* even reaches a higher value at the temperature of 500 K. At this temperature, $ZT_{max}$ is as high as 3.55 at the chemical energy $\mu = -1.21$ eV. Even at lower chemical energies, the figure of merit still shows a significant peak value $ZT_{max} = 1.66$ at $\mu = -0.24$ eV and $ZT_{max} = 1.82$ at $\mu = -0.25$ eV at the temperatures of 300 K and 500 K, respectively.

We show in figure 3(b) the thermoelectric performance of a larger GNF of cross-section $M_A = 6 \ (d_A = 1.136 \text{ nm})$ and $M_Z = 10 \ (d_Z = 1.1068 \text{ nm})$. The maximum figures of merit $ZT_{max} = 0.83$ and 1.11 are found at $\mu = 0.76$ eV and 0.79 eV at the temperatures of 300 and 500 K, respectively. Thus, this larger structure still provides high thermoelectric performance with $ZT > 1$.

It is also worth to note from the energy-position of the highest peak of $ZT$ that the structure of cross-section [$M_A = 4, M_Z = 5$] is a p-type thermoelectric material, while the structure of cross-section [$M_A = 6, M_Z = 10$] belongs to the n-type class.

In figure 3(c), the temperature dependence of $ZT$ is further investigated. Both structures have the same behavior as $ZT$ increases when increasing the temperature and reaches a peak value at a temperature $T_c$ before falling to lower values. This decreasing behavior at high temperature is actually due to the rapid increase of the electron thermal conductance $K_e$ as the temperature rises (not shown). In a previous work on GNRs [20], we showed that $K_e$ increases almost exponentially with temperature. Although at low and room temperature, $K_e$ is usually much smaller than $K_p$, at high temperature, $K_e$ becomes large and even comparable to $K_p$. When $K_e$ becomes predominant, the total thermal conductance $K = K_e + K_p$ increases and $ZT$ decreases. This behavior should hold for almost all graphene- and graphite-based structures.

Here we see that $ZT_{max}$ reaches 3.56 at 485 K in the structure [$M_A = 4, M_Z = 5$] and 1.12 at 505 K in the structure [$M_A = 6, M_Z = 10$].

*Additional enhancement by isotope doping*

Although the above results have revealed the excellent thermoelectric performance of GNFs, we show here that it is possible to further enhance their thermoelectric performance by introducing a certain form of defects in the lattice of carbon atoms.

To avoid the reduction of the electrical conductance usually associated with the presence of any kind of lattice defect, which might result in a lower thermal power factor, we chose to introduce a given amount of substitutional $^{14}$C isotopes into the GNFs initially made of pure $^{12}$C. This kind of doping was demonstrated to reduce significantly the in-plane phonon thermal conductance in graphene while maintaining the electronic properties unaltered[20,21], which is particularly relevant for thermoelectrics.

For each studied device, five random distributions with the same density of isotope doping were considered and the average value of the transmissions was considered to further study the phonon transport properties.

In figure 4(a), we show the average phonon transmission in the structure $[M_A = 4, M_Z = 5]$ for two active device lengths of 50 unit cells (32.8 nm) and 200 unit cells (131.2 nm), respectively, and for the isotope doping density of 50%. The result obtained for the structure $[M_A = 6, M_Z = 10]$ is given in the Supporting Information. As it can be seen, the phonon transmission is significantly reduced in the frequency-range between 100 cm$^{-1}$ and 200 cm$^{-1}$ compared to that of the pure $^{12}$C structure shown in figure 2(e). Moreover, it indicates that the phonon transmission is suppressed more strongly at higher frequencies and the impact of isotope doping is stronger in longer devices. This behavior is in agreement with what has been observed in previous studies of the in-plane transport in graphene nano-ribbons.[20]

As a result, the phonon thermal conductance obtained in figure 4(b) is obviously smaller than in the case without isotope doping (see figure 2(f)), i.e, $K_p = 0.063$ nW/K and 0.057 nW/K at room temperature in the doped structures of length $L_A = 32.8$ nm and 131.2 nm, respectively. The thermal conductance is thus about 36% and 42% smaller, respectively, than that obtained in the free-defect structure.

Thanks to the strong drop in the phonon thermal conductance, the figure of merit was found to be remarkably enhanced. In figure 4(c), *ZT* is plotted as a function of the chemical energy for both structure lengths at 300 K. The maximum of *ZT* is about 3.94 in the shorter device and about 4.2 in the longer device, i.e. significantly higher than the value of 2.85 obtained for the structure without isotope doping. As revealed in figure 3(c), *ZT* is possibly higher at higher temperatures. We therefore examined the temperature-dependence of *ZT* in these doped structures. It can be seen indeed in figure 4(d) that *ZT$_{max}$* reaches up to 4.79 and 5.07 for the lengths of 32.8 nm and 131.2 nm, respectively, at a temperature around $T_c = 450$ K.

We also investigated the impact of isotopes in the structure with a larger cross-section area [$M_A = 6, M_Z = 10$]. The result is shown in dotted-yellow line in figure 4(d). With 50% isotope doping density in the device of length $L_A = 131.2$ nm, the highest *ZT* obtained is about 1.95, which is 74% higher than without doping.

*Estimating the effect of inelastic scattering*

It has been demonstrated that in graphite layers (or in graphene sheets), the in-plane phonon mean free path (MFP) at room temperature is about 600-775 nm. [22,23] However, along the c-axis that is perpendicular to the graphite layers, the phonon MFP is much shorter and lies in the range 100-200 nm. [24,25] It is thus in the same order as the device length studied here, which makes it necessary to include the effect of inelastic phonon scattering, even if the electron transport is still considered to be ballistic.

In this section, we discuss the impact of inelastic scattering that might influence the thermoelectric performance of GNFs. It is known that to estimate the effect of inelastic scattering at device lengths comparable with the MFP, the mixed ballistic-diffusive nature of heat transport can be phenomenologically described through the formula of transmission written in the following form [21,26]

$$T = \frac{T_{\text{ballistic}}}{1 + L_A / MFP}. \qquad (1)$$

We considered the MFP values of 145.9 nm at 300 K and 68.4 nm at 500 K as predicted from molecular dynamics simulations[24] and the ballistic transmission $T_{\text{ballistic}}$ extracted from results of figures 2(e) and 4(a). If the MFP is assumed to be frequency-independent, for the device of length 131.2 nm, we have $T = 0.5265 \times T_{\text{ballistic}}$ at 300 K and $T = 0.3427 \times T_{\text{ballistic}}$ at 500 K.

Based on this estimation, the phonon conductance and the figure of merit *ZT* including inelastic phonon scattering can be easily deduced. We obtained $ZT_{max}$ = 4.52 and 6.56 at 300 K and 500 K, respectively, for the structure $[M_A = 4, M_Z = 5]$ as displayed in red opened symbols in figure 4(d). In the presence of 50% isotope doping in this structure, *ZT* even reaches 6.39 at 300 K and 8.09 at 500 K with scattering effects, as shown by red filled symbols in the same figure.

In the larger structure $[M_A = 6, M_Z = 10]$, $ZT_{max}$ is also significantly enhanced by phonon scattering, as shown by yellow opened symbols of figure 4(d) with $ZT_{max}$ = 1.43 and 2.18 at 300 K and 500 K, respectively, for the free-defect structure. In the $^{14}$C-doped structure (yellow filled symbols), we obtained $ZT_{max} = 2.59$ (at 300 K) and 3.02 (at 500 K).

## Conclusion

In conclusion, by means of atomistic simulation, we have demonstrated the outstanding thermoelectric capacity of GNFs with a figure of merit $ZT \sim 3.55$ observed in a fiber of diameter about 0.5 nm and with $ZT \sim 1.1$ for a 1.1 nm diameter GNF. Additionally, we have shown that by isotope doping engineering, $ZT$ can be further enhanced up to 5.07 in the narrow structure and up to 1.95 in the larger one. If inelastic phonon scattering is taken into account, the phonon conductance is further suppressed and becomes length dependent. For a device of 131 nm active length, the figure of merit in the narrow (larger) structure reaches 6.39 (2.50) at 300 K and 8.09 (3.02) at 500 K. The unveiled results are unprecedented for GNFs. They not only open perspectives for new applications of GNFs but also reveal GNFs as one of the most remarkable candidates for thermoelectric applications at the nanoscale.

## Methods

The electronic properties of GNFs were investigated using the density functional theory (DFT) with the SIESTA package.[16] Meanwhile, for the sake of convenience when introducing defects, a semi-empirical FC model was adopted for studying the phonon behavior and simulations were performed using our house-made code. Both DFT and FC models were coupled with Green's function formalism to explore the transport properties of GNFs. The details of the methods and the additional results can be found in the Supporting Information.

## Acknowledgments

This work was supported by the TRANSFLEXTEG Project of the H2020 program of the European Union under the reference 645241.

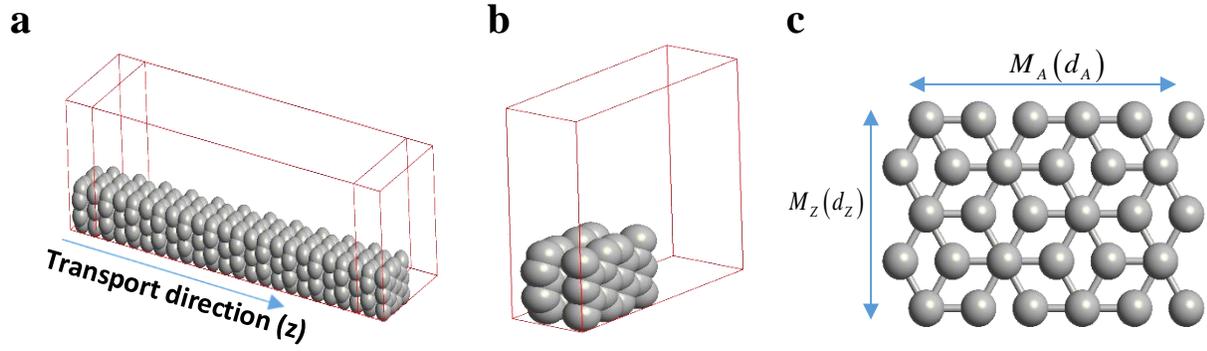

Figure 1. (a) Sketch of a nano device made of a "platelet" graphite nanofiber with transport along the *z* direction corresponding to the c-axis of graphite. (b) Typical unit cell of a graphite nanofiber. (c) Atomistic view of the cross section of a graphite nanofiber.

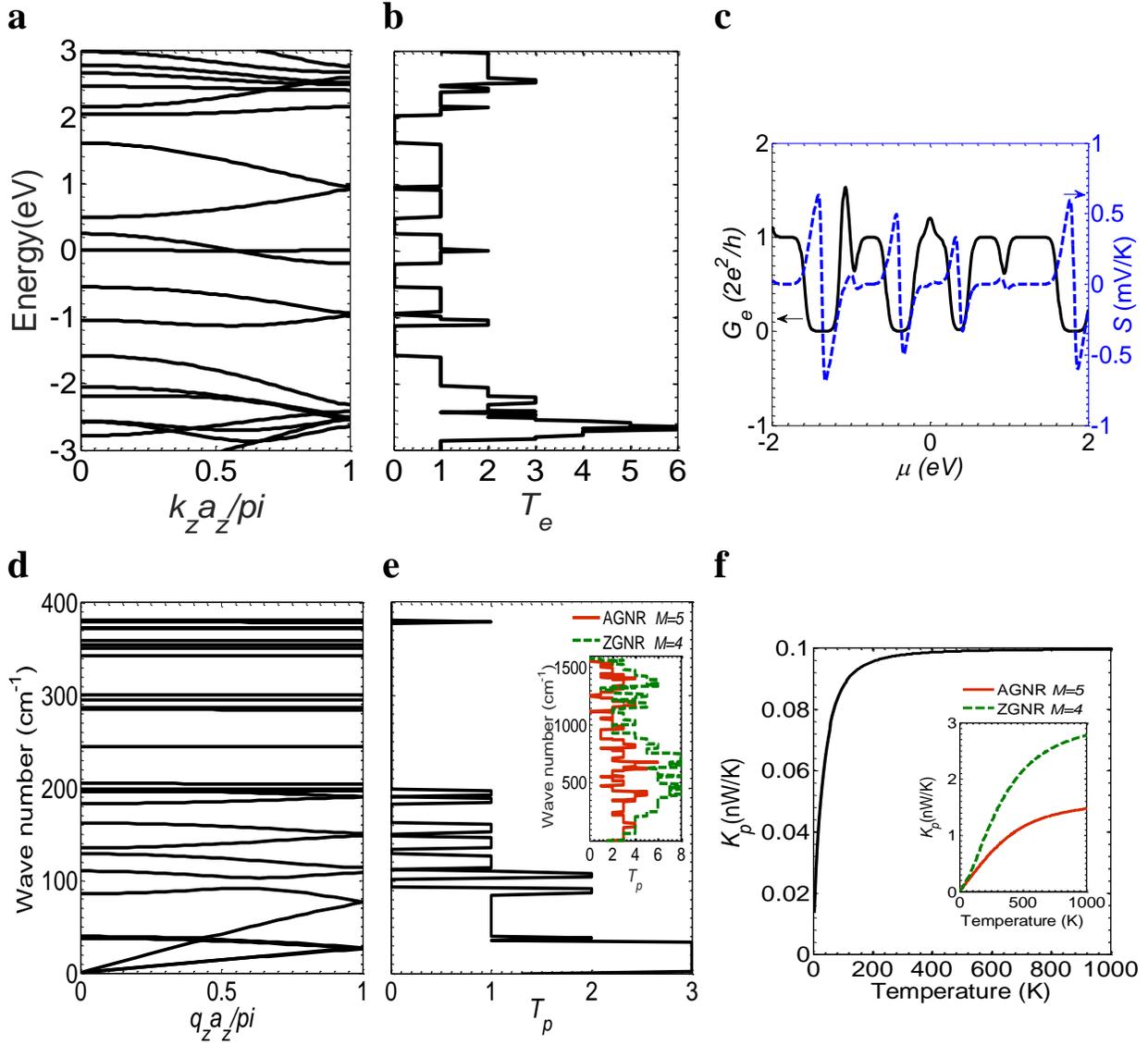

Figure 2. (a) Energy bands, (b) Electron transmission and (c) Electrical conductance and Seebeck coefficient at room temperature calculated by DFT. (d) Phonon bands, (e) Phonon transmission and (f) phonon thermal conductance calculated from FC model and Green's function formalism. All these results were obtained for graphite nanofiber [$M_A = 4, M_Z = 5$]. In figures (e) and (f), the insets show results obtained for in-plane transport in armchair and zigzag graphene ribbons for the sake of comparison.

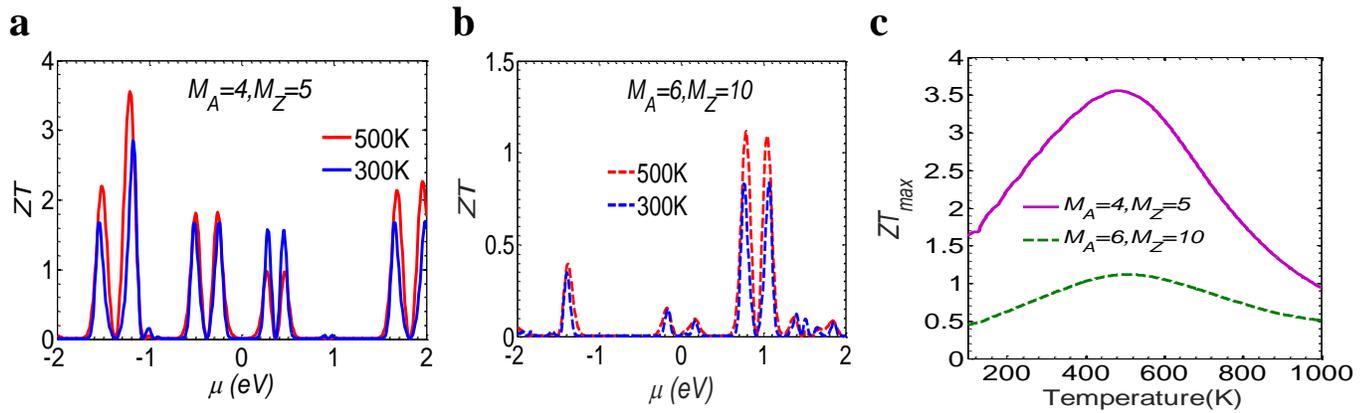

Figure 3. Figure of merit *ZT* as a function of chemical energy at different temperatures for pure graphite nanofiber of cross-section (a) $M_A = 4, M_Z = 5$ and (b) $M_A = 6, M_Z = 10$. (c) Maximum value $ZT_{max}$ as a function of temperature in both structures.

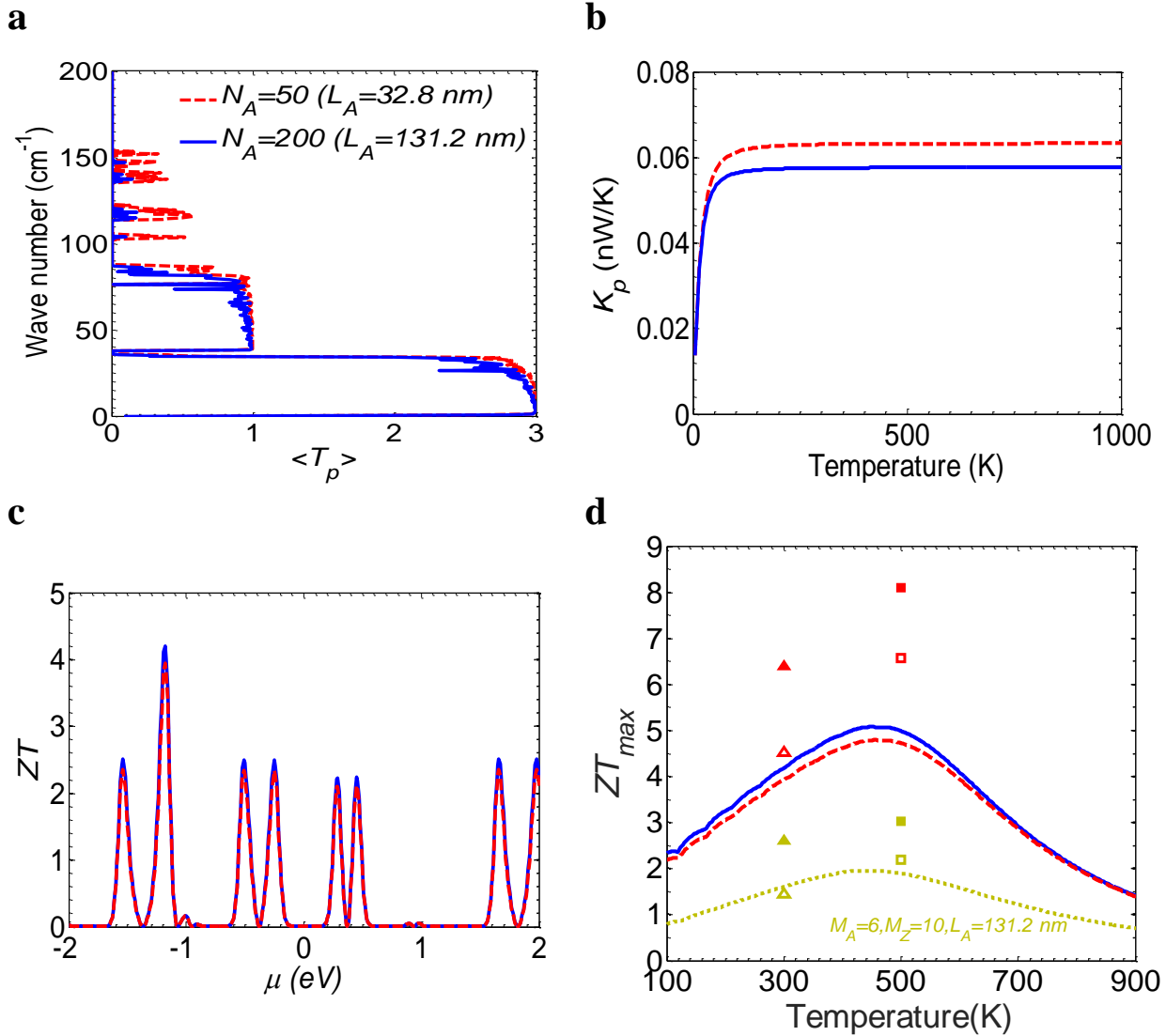

Figure 4. (a) Average phonon transmission for two different device lengths with 50% of $C^{14}$ isotope doping. (b) Phonon conductance calculated as a function of temperature for the two defected structures. (c) $ZT$ at room temperature and (d) $ZT_{max}$ as a function of temperature in the presence of isotope doping. Except the results plotted in the dotted-yellow line in (d) that was obtained for the structure [$M_A = 6, M_Z = 10$], all results were obtained for the structure [$M_A = 4, M_Z = 5$]. Open and filled symbols are results including the inelastic phonon scattering for structures without and with isotope doping, respectively.

# High thermoelectric performance of graphite nanofibers

Van-Truong Tran[1*], Jérôme Saint-Martin[2], Philippe Dollfus[2] and Sebastian Volz[1, 3]

[1]EM2C, CentraleSupélec, Université Paris Saclay, CNRS, 92295 Châtenay Malabry, France
[2]C2N, CNRS, Université Paris-sud, Université Paris Saclay, 91405 Orsay, France
[3]LIMMS, Institute of Industrial Science, University of Tokyo, CNRS-IIS UMI2820, 4-6-1 Komaba Meguro-Ku, Tokyo 153-8505, Japan
*vantruongtran.nanophys@gmail.com

**Supporting information for methods**

*Density functional theory (DFT) calculations*

In this work, we used the localized orbital density functional theory implemented within the SIESTA package.[1] To include the van der Waals (vdW) interactions between atoms in different layers, the pseudo-potentials were corrected by adding the DRSLL-vdW interactions[2] and generated with Atom program[3] under Troullier-Martins scheme.

All DFT calculations were performed after connecting hydrogen atoms to all edge-carbon atoms to compensate the sp2 dangling bonds[4] and thus to avoid strong edge deformation which could generate unexpected states in the gap due to the charge transfer effect. This is also consistent with the treatment of the Force Constant or Tight Binding model herein the truncation at the edges is usually considered with extended lines of carbon atoms on each side of the edges to compensate the sp2 dangling bonds of the edge carbon atoms, and these extended lines are treated as hard walls that do not impact on the results of tight binding calculations.

Additionally, in all structures, lattice vectors along *x*, *y* directions (perpendicular to the fiber axis) were set sufficient large (40 Angstrom) to avoid interactions between the system and its images because of the periodic boundary condition.

In all DFT calculations, the double Zeta polarized (DZP) orbital basis set was used and a mesh energy cutoff of 400 Ry was taken. A Monkhorst-Pack[5] 1×1×10 was chosen for the relax calculations, while for the band structure and transport studies a grid 1×1×20 was adopted. All structures were relaxed within conjugate-gradients (CG) method until the total force was less than 0.05 eV/Angstrom. The variable cell in relaxation was also set up to search for an appropriate equilibrium distance between graphite layers.

*Force Constant (FC) model*

For the study of phonons, we employed a Force Constant (FC) model in which the secular equation for phonons is written:

$$D\mathbb{U} = \omega^2 \mathbb{U}, \tag{S1}$$

where $\mathbb{U}$ is the column matrix containing the amplitude vectors of vibration at all lattice sites and $\omega$ is the angular frequency, $D$ is the Dynamical matrix which is calculated as [6,7]

$$D = \left[ D_{3\times 3}^{ij} \right] = \begin{bmatrix} \begin{cases} -\dfrac{K_{ij}}{\sqrt{M_i M_j}} & \text{for } j \neq i \\ \dfrac{\sum_{n \neq i} K_{in}}{M_i} & \text{for } j = i \end{cases} \end{bmatrix} \tag{S2}$$

The coupling tensor $K_{ij}$ between the *i*-th and *j*-th atoms is defined depending on whether the type of interaction is in-plane or inter-plane, i.e.

(i) For in-plane interactions, $K_{ij}$ is determined by a unitary in-plane rotation [6,7]

$$K_{ij} = U^{-1}(\theta_{ij}) K^0{}_{ij} U(\theta_{ij}) \tag{S3}$$

where

$$U(\theta_{ij}) = \begin{bmatrix} \cos(\theta_{ij}) & \sin(\theta_{ij}) & 0 \\ -\sin(\theta_{ij}) & \cos(\theta_{ij}) & 0 \\ 0 & 0 & 1 \end{bmatrix} \tag{S4}$$

is the rotation matrix [7] and $\theta_{ij}$ is the anticlockwise rotating angle formed between the the *x*-axis and the vector joining the *i*-th to the *j*-th atoms. In equation (S3), $K^0_{ij}$ is the force constant tensor that contains the force constant parameters [7]

$$K^0_{ij} = \begin{pmatrix} \Phi_r & 0 & 0 \\ 0 & \Phi_{t_i} & 0 \\ 0 & 0 & \Phi_{t_o} \end{pmatrix} \quad (S5)$$

where $\Phi_r, \Phi_{t_i}$, and $\Phi_{t_o}$ are the force constant coupling parameters in the radial, in-plane and out-of-plane directions, respectively, and their values usually decay with the neighboring distance. In this work, a four nearest neighbor range was considered and thus twelve parameters for in-plane coupling was taken from Ref.[8].

(ii) For the vdW or interlayer interactions, we employed the spherically symmetric interatomic potential model, in which each component of the coupling tensor $K_{ij}$ is defined by:[9]

$$\left(K_{ij}\right)_{kk'} = \delta\left(r^{ij}\right) \cdot \frac{r^{ij}_k \cdot r^{ij}_{k'}}{\left(r^{ij}\right)^2} \quad (S5)$$

with $k, k' = \overline{1,3}$ or $x, y, z$. $r^{ij}$ is the vector joining the *i*-th to the *j*-th atoms and $\delta\left(r^{ij}\right)$ is the decaying component $\delta\left(r^{ij}\right) = A.\exp\left(-r^{ij}/B\right)$ with empirical parameters $A = 573.76$ N/m, $B = 0.05$ nm. It should be noted that equation (S5) does not contain the minus sign "-" as in ref. [9] because this sign has been included in equation (S2). Moreover, to have the best fit between the FC model and the experimental data for bulk graphite, in the FC model the distance between two graphite layers was taken equal to 0.328 nm.

### *Green's function formalism for transport study*

To study the transport properties of both electrons and phonons, the atomistic Green' function formalism was employed.[10] All device structures were divided into three parts: the left and right leads and the device region (central region). The leads were treated as semi-infinite regions while the central region is a finite region containing the left lead extension, the active region, and the right lead extension. In our calculations, the left (right) lead extension was chosen of the size of one unit cell (6.56 Angstrom) that is enough to make the left (right) lead isolated from the active region. The active (or scattering) region contains $N_A$ unit cells with the length $L_A = N_A \times a_z$ where $a_z \approx 0.656$ nm.

The Hamiltonian $H$ or the Dynamical matrix and the overlap matrix $S$ of the whole device structure were split into three parts $H_L$, $H_D$, $H_R$ and $S_L$, $S_D$, $S_R$ (similarly $D_L$, $D_D$, $D_R$ for phonons) as the Hamiltonians and overlap matrices of the left lead, device part(central region) and right lead, respectively, including the couplings between the device and the two leads $H_{DL}$, $H_{DR}$, $S_{DL}$, and $S_{DR}$ ($D_{DL}$, $D_{DR}$ for phonons). The Green's function of the device region is defined by the equation

$$G = \left[ (E + i.\eta).S_D - H_D - \Sigma^s_L - \Sigma^s_R \right]^{-1}, \tag{S6}$$

where $\eta$ is a positive infinitesimal number and

$$\begin{aligned}\Sigma^s_L &= \left( E^+.S_{DL} - H_{DL} \right) G^0_L \left( E^+.S_{LD} - H_{LD} \right) \\ \Sigma^s_R &= \left( E^+.S_{DR} - H_{DR} \right) G^0_R \left( E^+.S_{RD} - H_{RD} \right)\end{aligned} \tag{S7}$$

are the surface self-energies describing the energy-dependent coupling with the left and right leads. $G^0_{L(R)}$ is the surface Green's function of the isolated left (right) lead.[11,12]

For phonons, we just need to replace energy $E$ by $\omega^2$, and $H_D$, $H_{DL}$, $H_{LD}$, $H_{DR}$, $H_{RD}$ by $D_D$, $D_{DL}$, $D_{LD}$, $D_{DR}$, $D_{RD}$, respectively. We also set $S_D = \mathbf{1}, S_{DL} = S_{LD} = S_{DR} = S_{RD} = \mathbf{0}$ for phonons.

The size of the device Green's function was reduced by making use of the recursive technique.[13,14] The electron (phonon) transmission was computed as [15]

$$T_{e(p)} = Trace \left\{ \Gamma^s_L \left[ i \left( G_{11} - G_{11}^\dagger \right) - G_{11} \Gamma^s_L G_{11}^\dagger \right] \right\}, \tag{S8}$$

where $\Gamma^s_{L(R)} = i \left( \Sigma^s_{L(R)} - \Sigma^{s\;\dagger}_{L(R)} \right)$ denotes the surface injection rate at the left (right) lead.

The electrical conductance, the Seebeck coefficient and the electron thermal conductance were computed within the Landauer-Onsager's approach, i.e,[16]

$$G_e(\mu,T) = e^2 . L_0(\mu,T)$$
$$S(\mu,T) = \frac{1}{eT} \cdot \frac{L_1(\mu,T)}{L_0(\mu,T)} \quad \text{(S9)}$$
$$\kappa_e(\mu,T) = \frac{1}{T} \cdot \left[ L_2(\mu,T) - \frac{L_1(\mu,T)^2}{L_0(\mu,T)} \right]$$

where the intermediate functions $L_n$ may be written in the form [6,15,17]

$$L_n(\mu,T) = \frac{1}{h} \int_{-\infty}^{+\infty} dE . T_e(E) . (2K_b T)^{n-1} . g^e_n(E,\mu,T), \quad \text{(S10)}$$

where $g^e_n(E,\mu,T) = \left( \frac{E-\mu}{2K_b T} \right)^n / \cosh^2\left( \frac{E-\mu}{2K_b T} \right)$ is a dimensionless function and $K_b$ is the Boltzmann constant.

Similarly, the Landauer-like formula was used to compute the phonon thermal conductance [15,16]

$$K_p = \frac{K_b}{2\pi} \int_0^{\infty} d\omega . T_p(\omega) . g^p(\omega,T), \quad \text{(S10)}$$

where $g^p(\omega,T) = \left( \frac{\hbar\omega}{2K_b T} \right)^2 / \sinh^2\left( \frac{\hbar\omega}{2K_b T} \right)$ is also a dimensionless function, as $g^e_n(E,\mu,T)$.

Finally, to assess the thermoelectric ability of a structure, the figure of merit $ZT$ is used as the essential criterion and is calculated as [6,18,19]

$$ZT = \frac{G_e . S^2}{K_e + K_p} . T \quad \text{(S10)}$$

The electronic relaxation and band structure calculations were performed within the SIESTA module while the electron transmission was implemented with TransSIESTA module of the SIESTA package. The Force Constant model and Green's function technique for phonons were treated within our house-made code. We also used the Virtual Nanolab (VNL) [20] as a graphic user interface for the SIESTA code.

**Supporting information for additional results**

To validate the FC model, we performed phonon dispersion calculations and compared the results to the experimental data of ref. [21]. Since the direction of interest in this work is along the c-axis, we focussed on the phonon bands along the GA k-path (see figure S1(b)). As it can be seen in figure S1(c), the solid black lines obtained from the FC model are in excellent agreement with the experimental data, demonstrating the quality of this model.

Remarkably, the phonon frequency range along the c-axis is much shorter than along other axes [21], confirming that the vdW interaction between graphite layers is very weak.

In figure S2, the electronic and phononic properties of the GNF of cross-section size [$M_A = 6$, $M_Z = 10$] are displayed. As it can be observed in figure S2(a), in the energy range from -3 eV to 3 eV, only one visible gap is found around 1 eV. Accordingly, a single-electron transmission gap and a large Seebeck coefficient are obtained around this energy, as can be seen in figures S2(b) and S2(c), respectively. Interestingly, compared to the phonon dispersion of the small cross-section GNF [$M_A = 4, M_Z = 5$] presented in figure 2(d) in the main article, the phonon dispersion shown in figure S2(d) indicates that there are more dispersive phonon bands at higher frequencies as the size of the cross-section increases. Although the group velocities of these high-frequency bands are low compared to that of the acoustic bands, they add a contribution to the transmission, which leads to a larger phonon thermal conductance (see figures S2(e) and S2(f)). The higher thermal conductance usually leads to lower thermoelectric performance. This explains the fact that figure of merit obtained in the macroscopic graphite fiber as studied in ref. [22] is very low.

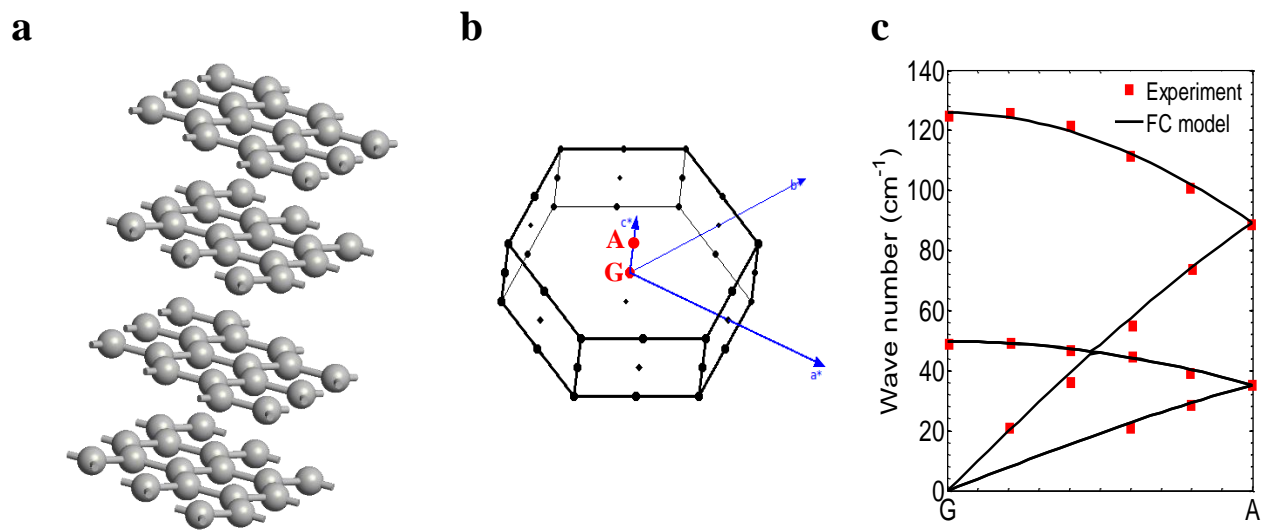

Figure S1. (a) Example of 3x3x2 unit cells of bulk graphite. (b) Brillouin zone of bulk graphite, generated using the Xcrysden software.[23] (c) Validation of the FC model for graphite by comparison with the experimental data of ref. [21].

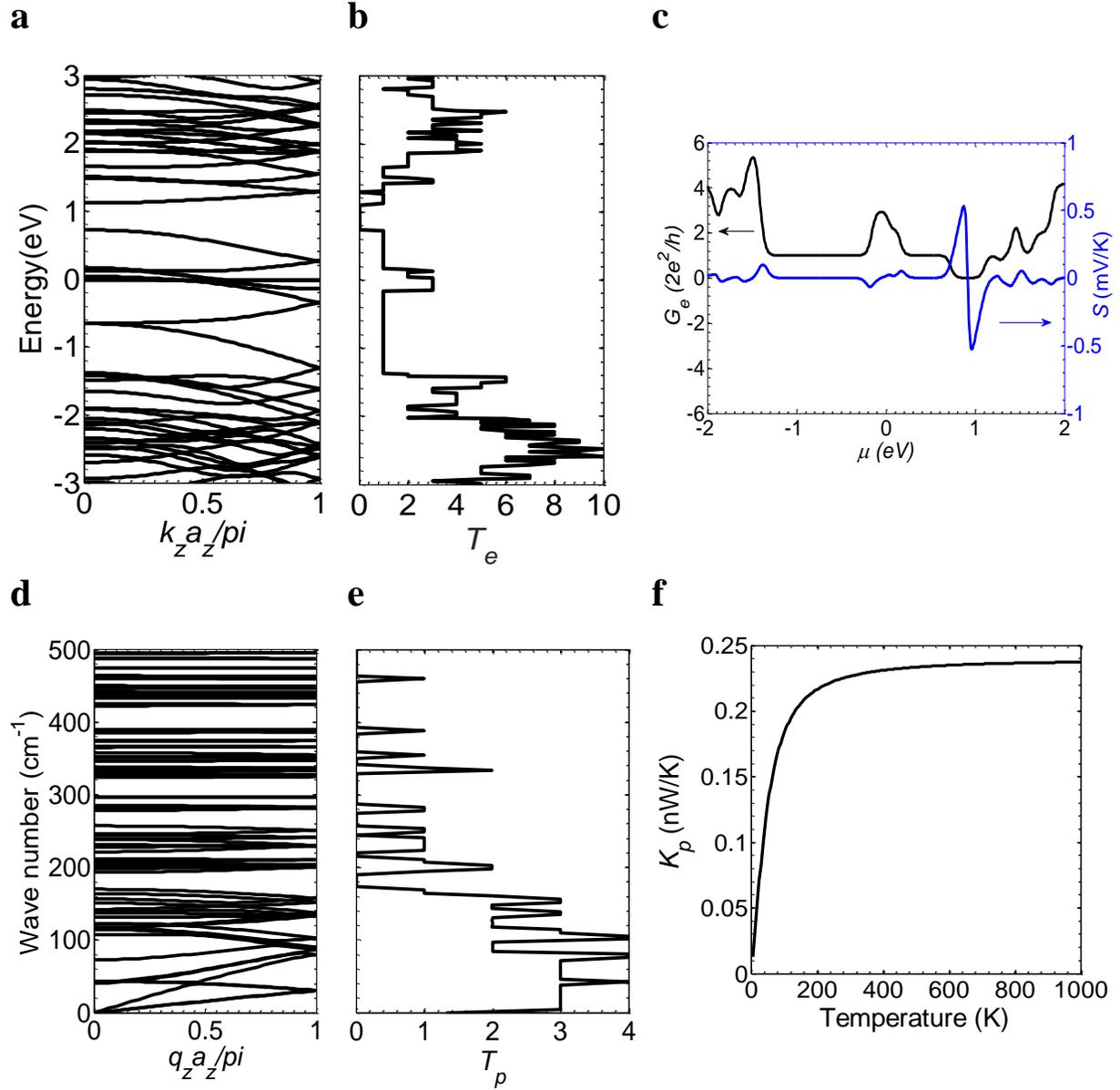

Figure S2. GNF of cross-section size [ $M_A = 6, M_Z = 10$ ]. (a) Energy bands, (b) Electron transmission and (c) Electrical conductance and Seebeck coefficient at room temperature calculated by DFT. (d) Phonon dispersion, (e) Phonon transmission and (f) Phonon thermal conductance calculated from FC model and Green's function formalism.